\documentclass[conference]{IEEEtran}
\IEEEoverridecommandlockouts

\usepackage{cite}
\usepackage{amsmath,amssymb,amsfonts}
\usepackage{algorithmic}
\usepackage{graphicx}
\usepackage{textcomp}
\usepackage{xcolor}

\usepackage{array}
\usepackage{booktabs}
\usepackage{multirow}
\usepackage{epsfig}
\usepackage{bm}
\usepackage{algorithm}
\usepackage{diagbox}
\usepackage{amssymb}
\usepackage{makecell}
\usepackage{adjustbox}
\usepackage{enumitem}
\usepackage{marvosym}
\usepackage{braket}
\usepackage{color}
\def\BibTeX{{\rm B\kern-.05em{\sc i\kern-.025em b}\kern-.08em
    T\kern-.1667em\lower.7ex\hbox{E}\kern-.125emX}}

\begin{document}

\title{Exploring Audio-Visual Information Fusion for Sound Event Localization and Detection In Low-Resource Realistic Scenarios}

\author{
\IEEEauthorblockN{1\textsuperscript{st} Ya Jiang}
\IEEEauthorblockA{
\textit{University of Science and} \\ \textit{Technology of China}\\
Hefei, China \\}
\and
\IEEEauthorblockN{2\textsuperscript{nd} Qing Wang\textsuperscript{*}
\thanks{\textsuperscript{*}corresponding author}}
\IEEEauthorblockA{
\textit{University of Science and} \\ \textit{Technology of China}\\
Hefei, China \\}
\and
\IEEEauthorblockN{3\textsuperscript{rd} Jun Du}
\IEEEauthorblockA{
\textit{University of Science and} \\ \textit{Technology of China}\\
Hefei, China \\}

\and
\IEEEauthorblockN{4\textsuperscript{th} Maocheng Hu}
\IEEEauthorblockA{
\textit{National Intelligent Voice}  \\ \textit{ Innovation Center} \\
Hefei, China \\}
\and
\\
\IEEEauthorblockN{\quad 5\textsuperscript{th} Pengfei Hu}
\IEEEauthorblockA{
\textit{\quad \quad University of Science and} \\ \textit{\quad \quad Technology of China}\\
\quad \quad Hefei, China \\}
\and
\\
\IEEEauthorblockN{6\textsuperscript{th} Zeyan Liu}
\IEEEauthorblockA{
\textit{University of Science and} \\ \textit{Technology of China}\\
Hefei, China \\}
\and
\and
\\
\IEEEauthorblockN{7\textsuperscript{th} Shi Cheng}
\IEEEauthorblockA{
\textit{University of Science and} \\ \textit{Technology of China}\\
Hefei, China \\}
\and
\\
\IEEEauthorblockN{8\textsuperscript{th} Zhaoxu Nian}
\IEEEauthorblockA{
\textit{University of Science and} \\ \textit{Technology of China}\\
Hefei, China \\}
\and
\\
\IEEEauthorblockN{\quad 9\textsuperscript{th} Yuxuan Dong}
\IEEEauthorblockA{
\textit{\quad \quad University of Science and} \\ \textit{\quad\quad Technology of China}\\
\quad\quad Hefei, China \\}
\and
\and
\\
\IEEEauthorblockN{\quad 10\textsuperscript{th} Mingqi Cai}
\IEEEauthorblockA{\textit{\quad\quad iFlytek Research} \\
\quad\quad Hefei, China \\}
\and
\and
\and
\and
\and
\and
\\
\IEEEauthorblockN{11\textsuperscript{th} Xin Fang}
\IEEEauthorblockA{\textit{iFlytek Research} \\
Hefei, China \\}
\and
\and
\and
\and
\and
\\
\IEEEauthorblockN{ 12\textsuperscript{th} Chin-Hui Lee}
\IEEEauthorblockA{\textit{Georgia Institute of } \\ \textit{Technology} \\
Atlanta, USA \\}
}

\maketitle

\begin{abstract}
This study presents an audio-visual information fusion approach to sound event localization and detection (SELD) in low-resource scenarios. We aim at utilizing audio and video modality information through cross-modal learning and multi-modal fusion. First, we propose a cross-modal teacher-student learning (TSL) framework to transfer information from an audio-only teacher model, trained on a rich collection of audio data with multiple data augmentation techniques, to an audio-visual student model trained with only a limited set of multi-modal data. Next, we propose a two-stage audio-visual fusion strategy, consisting of an early feature fusion and a late video-guided decision fusion to exploit synergies between audio and video modalities. Finally, we introduce an innovative video pixel swapping (VPS) technique to extend an audio channel swapping (ACS) method to an audio-visual joint augmentation. Evaluation results on the Detection and Classification of Acoustic Scenes and Events (DCASE) 2023 Challenge data set demonstrate significant improvements in SELD performances. Furthermore, our submission to the SELD task of the DCASE 2023 Challenge ranks first place by effectively integrating the proposed techniques into a model ensemble.
\end{abstract}


\section{Introduction}
\label{sec:intro}
The goal of sound event localization and detection (SELD) is to detect the onsets and offsets of occurrences of specific sound events over time, while simultaneously tracking their spatial positions during the active period. The spatiotemporal information acquired from SELD systems holds broad applications in different audio-visual applications, such as self-localization, smart home, and audio surveillance \cite{foggia2015audio}.

SELD usually consists of two subtasks: sound event detection (SED) and sound source localization (SSL) which primarily focuses on direction-of-arrival (DOA) estimation in this paper. The SED task typically employs various classification methods such as Gaussian mixture models (GMMs) \cite{heittola2013context}, hidden Markov models (HMMs) \cite{butko2011two} and recurrent neural networks (RNNs) \cite{hayashi2017duration} to identify the temporal occurrences of each sound event of interest. DOA estimation methods generally include parametric-based approaches, such as subspace-based methods \cite{schmidt1986multiple}, time-difference-of-arrival (TDOA) techniques \cite{knapp1976generalized}, signal synchronization-based approaches \cite{dmochowski2007generalized} and data-driven neural-network(NN)-based approaches \cite{sundar2020raw} \cite{nguyen2020robust}. Recently, there has been a considerable amount of attention on jointly performing SED and DOA estimation, as introduced in the Detection and Classification of Acoustic Scenes and Events (DCASE) 2019 Challenge \cite{politis2020overview}. SELDnet \cite{adavanne2018sound} was then proposed to utilize two parallel branches for SED and DOA estimation, respectively. Afterward, various NN-based solutions for SELD have emerged. In \cite{shimada2021accdoa}, Shimada \emph{et al.} proposed an activity-coupled Cartesian DOA (ACCDOA) and expanded it to multi-ACCDOA \cite{shimada2022multi} to distinguish multiple overlapping sound events of the same category. Currently, the majority of SELD networks process multi-channel audio inputs.

Considering that humans perceive multi-modal cues to explore real-world events, some studies have suggested integrating audio and video modalities to gain better insight into the SELD problem. However, most audio-visual detection and localization studies primarily focus on localization within video frames. This has led to the development of multiple techniques \cite{tian2018audio, xuan2020cross}, represented by the publicly available audio-visual event localization (AVE) task. These researches concentrate on categorizing events occurring in each video segment, as well as locating event boundaries based on a given visual or auditory query, where an audio-visual event is both audible and visible in the video. Nonetheless, it is challenging to localize a visual object emitting sound in spatial positions, because audio operates in a 3D space while video operates on a 2D image plane. Some studies proposed transformations between a spatial DOA and a target image location using calibrated sensors\cite{liu2019audio, qian2019multi}. Specifically, \cite{qian2022audio} attempted to localize and track speaker DOAs using an audio-visual cross-modal attentive fusion framework. \cite{jiang2022egocentric} proposed to localize active speakers in a full 360$^\circ$ field of view.

Recently, the DCASE 2023 Challenge introduced an audio-visual track utilizing synchronized audio and video recordings. Although video data can provide valuable cues to mitigate the challenges in characterizing the spatiotemporal features of acoustic scenes, the availability of real audio-visual data is extremely limited, in this case with merely 3.83 hours, further adding to the challenges of effectively utilizing the information from video modality for solving the SELD problem. Human faces are blurred due to privacy concerns, making it impossible to extract speech activities from the video. Additionally, the information embedded in video frames is too redundant, leading to the potential duplication of sound targets. Therefore, we propose the use of cross-modal transfer learning followed by multi-modal fusions for audio-visual SELD (AV-SELD) in low-resource realistic scenarios. We highlight the three key contributions of this work as follows:

\begin{enumerate}[label=(\arabic*), itemsep=-4pt]
    \item We design a cross-modal teacher-student learning (TSL) framework to perform transfer learning from the teacher model trained on abundant external audio data to the student model with limited audio-visual data.
    \item We introduce a two-stage process employing feature fusion and video-guided decision fusion to further improve the localization precision in SELD models.
    \item We propose an efficient video pixel swapping (VPS) method to jointly augment multi-modal data eightfold and enhance the robustness of the student model.
\end{enumerate}

\begin{figure}[t]
	\centering
	\includegraphics[width=1\linewidth]{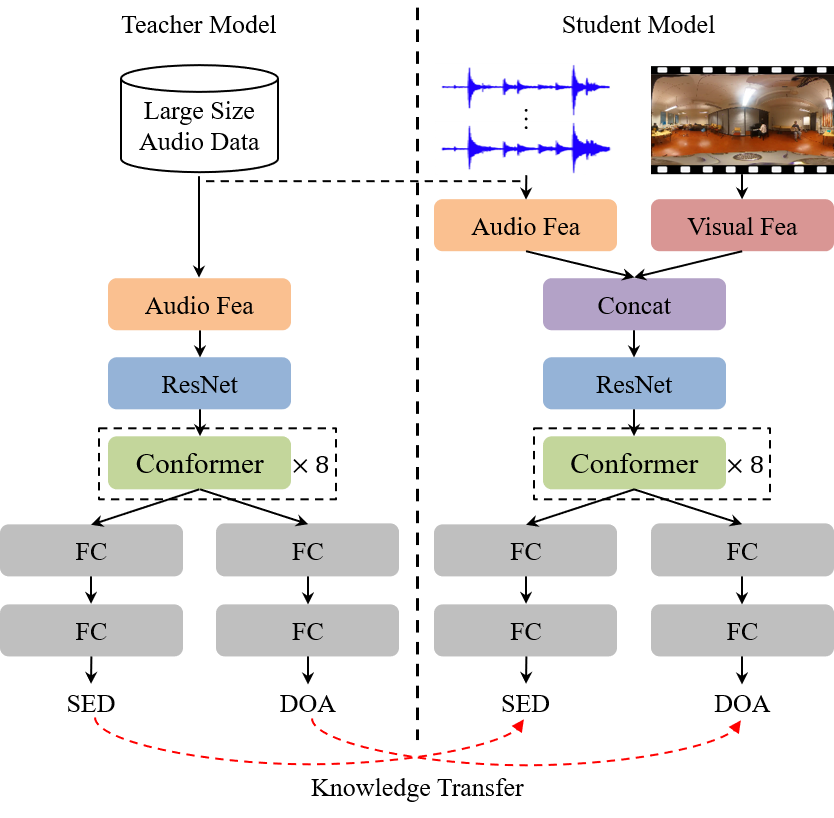}
	\caption{The proposed AV-SELD model is based on cross-modal teacher-student learning and multi-modal fusion, as shown in the overall architecture.}
    \label{fig: architecture}
\end{figure}

\section{PROPOSED METHOD for AV-SELD}
The overall flowchart of the proposed AV-SELD framework using cross-modal teacher-student learning is illustrated in Fig. \ref{fig: architecture}, consisting of the teacher model trained with extensive audio data and the student model built with limited audio-visual data. Besides, a two-stage audio-visual deep fusion strategy and joint audio-visual data augmentation are incorporated to further improve the SELD performance. The details are elaborated in the following subsections.

\subsection{Audio-only teacher model training}
\label{ssec: ao model}
The audio-only teacher model is constructed as the ResNet-Conformer (RC) architecture proposed in our previous work \cite{niu2023experimental} in the DCASE 2022 Challenge, utilizing an 18-layer ResNet network followed by an 8-layer Conformer module as illustrated in the left branch of Fig. \ref{fig: architecture}. We slice the 4-channel audio data into 10-second segments and extract 4-channel log Mel-spectrograms and 3-channel intensity vector (IV) features by applying short-term Fourier transform (STFT) with a frame length of 40 ms and a frame hop of 20 ms. Consequently, the concatenated 500$\times$7$\times$64 audio features are fed into the model. The ResNet-Conformer follows two parallel branches to perform SED and DOA estimation respectively, each containing two fully connected (FC) layers. The audio-only teacher network is trained using a multi-objective learning framework:
\begin{equation}
\label{eq: seld_loss}
    \mathcal{L}_{\mathrm{SELD}}^{\mathrm{M}}=\beta_1 \mathcal{L}_{\mathrm{SED}}^{\mathrm{M}}+\beta_2 \mathcal{L}_{\mathrm{DOA}}^{\mathrm{M}}
\end{equation}

where $\mathrm{M\in\{T, S, TS\}}$. When $\mathrm{M=T}$, $\mathcal{L}_{\mathrm{SELD}}^{\mathrm{T}}$ represents the loss function used in teacher model training. When $\mathrm{M=S}$, $\mathcal{L}_{\mathrm{SELD}}^{\mathrm{S}}$ denotes the original loss function used in student model training in Section \ref{ssec: av model}. When $\mathrm{M=TS}$, $\mathcal{L}_{\mathrm{SELD}}^{\mathrm{TS}}$ indicates the teacher-instructed loss function used in cross-modal teacher-student learning in Section \ref{ssec: cross-modal tsl}. $\beta_1=0.1$ and $\beta_2=1$ are the weighting factors set for SED loss $\mathcal{L}_{\mathrm{SED}}$ and DOA loss $\mathcal{L}_{\mathrm{DOA}}$ respectively. The SED and DOA subtasks are optimized by minimizing the binary cross entropy (BCE) and mean squared error (MSE) criteria respectively, as defined below:
\begin{align}
\mathcal{L}_{\mathrm{SED}}^{\mathrm{T}}= & -\frac{1}{K N} \sum_{k, n}\left[y_{k, n}^{\mathrm{T}} \log \hat{y}_{k, n}^{\mathrm{T}}\right. \\
& \left.+\left(1-y_{k, n}^{\mathrm{T}}\right) \log \left(1-\hat{y}_{k, n}^{\mathrm{T}}\right)\right] \nonumber \\ 
\mathcal{L}_{\mathrm{DOA}}^{\mathrm{T}}= & \frac{1}{K N} \sum_{k, n}\left\|\left(\hat{\mathbf{o}}_{k, n}^{\mathrm{T}}-\mathbf{o}_{k, n}^{\mathrm{T}}\right) y_{k, n}^{\mathrm{T}}\right\|^2
\end{align}where $\{y_{k, n}^{\mathrm{T}}, \hat{y}_{k, n}^{\mathrm{T}}\}$ and $\{\mathbf{o}_{k, n}^{\mathrm{T}}, \hat{\mathbf{o}}_{k, n}^{\mathrm{T}}\}$ represent the ground truth and model output for SED and DOA of the $n$-th sound event at the $k$-th frame, respectively. $\mathbf{o}_{k, n}^{\mathrm{T}}$ and $\hat{\mathbf{o}}_{k, n}^{\mathrm{T}}$ represent Cartesian position vectors. $K$ denotes the total number of frames in a batch and $N$ denotes the number of classes.

During the training of the teacher model, we incorporate external audio data with multiple data augmentation techniques to enhance the model's stability. Firstly, we leverage the additional 20 hours of simulated data provided by the DCASE 2023 Challenge and apply the ACS augmentation to generate extensive audio data, following the approach described in our previous work \cite{wang2020four}. The ACS augmentation involves performing transformations directly to four audio channels, resulting in eight different DOA representations and effectively increasing the amount of audio data eightfold. Additionally, we employ mixup \cite{zhang2018mixup} augmentation to enhance the diversity of the model input features.

\subsection{Audio-visual student model training}
\label{ssec: av model}
The student model is an audio-visual SELD network sharing the same backbone structure as the teacher model and incorporating the audio-visual feature fusion to use spatiotemporal information from multiple modalities. The audio features are extracted in the same way as the teacher model. For the visual modality, we select 10 video frames per second evenly to perform object detection using Faster-RCNN \cite{ren2015faster}, which can predict various classes of objects. Based on the detected boxes of the human class, a keypoint detection model, e.g., HRNet \cite{sun2019deep} pre-trained on COCO-WholeBody dataset \cite{jin2020whole}, then predicts five keypoints of visible speakers, namely mouths, left and right hands, and left and right feet for each corresponding image. We use the pixel coordinate of the mouth of each person to generate two Gaussian-like vectors, indicating likelihoods of speakers appearing along horizontal and vertical axes \cite{qian2022audio}. The center is the same as the mouth position and the standard deviation is proportional to a pre-defined width and height. We extract visual features for each person and pad with zeros in cases where fewer than six people are present simultaneously, resulting in a 100$\times$6$\times$2$\times$64 Gaussian feature vector for a 10-second clip. The visual features are replicated five times along the temporal dimension and reshaped to 500$\times$12$\times$64 to be concatenated with audio features, leading to 19-channel spatiotemporal representations. The fused multi-modal features are fed into ResNet-Conformer as shown in the right branch of Fig. \ref{fig: architecture}.

To address the problem of data sparsity, we propose a novel VPS method to extend ACS approach to audio-visual joint data augmentation, which matches multi-modal data in pairs. Given that the ACS method is inherently realized by swapping the four microphones arranged on a spherical baffle spatially, we can regard a $360^\circ$ panoramic video frame image as a cylindrical surface with the camera located at the center of the cylinder. Therefore, the pixel resolution of the image, e.g., 1920 $\times$ 960, corresponds to an azimuth angle range of $\phi \in [180^\circ, -180^\circ ]$ and an elevation angle range of $\theta \in [-90^\circ, 90^\circ]$. Based on the above analysis, we can design eight corresponding translations and inversions of pixel coordinates to jointly expand audio-visual data efficiently. Further details can be found in our technical report \cite{Du_NERCSLIP_task3_report}. Taking one transformation, $\phi=\phi+\pi, \theta=\theta$ for example, which means the original DOA azimuth angle $\phi$ is rotated by 180$^\circ$ while the elevation angle remains the same. Accordingly in the process of VPS, the horizontal pixel points are translated by 960 pixel points along the negative direction, while the vertical pixel points remain unchanged in the corresponding video image.

\subsection{Cross-modal teacher-student learning}
\label{ssec: cross-modal tsl}
We construct a cross-modal teacher-student learning (TSL) framework based on teacher weights transferring and cross-modal loss function. In the first step, pre-trained weights of the teacher model are utilized to initialize the first 7 channels of the student model’s input feature maps with the remaining 12 channels randomly initialized. Secondly, we design a teacher-instructed loss function $\mathcal{L}_{\mathrm{SELD}}^{\mathrm{TS}}$ as a regularization term of the original student loss function $\mathcal{L}_{\mathrm{SELD}}^{\mathrm{S}}$ calculated in the same way as Eq. \ref{eq: seld_loss}. The final TSL loss function is formulated as Eq. \ref{eq: TS-SELD}. Specifically, we employ Kullback-Leibler (KL) divergence to regularize SED loss inspired by \cite{zhou2021audio}, computed with the SED output of the teacher and student network as Eq. \ref{eq: TS SED}. Meanwhile, the DOA loss of the student model $\mathcal{L}_{\mathrm{DOA}}^{\mathrm{S}}$ is regularized using the outputs in the teacher model as Eq. \ref{eq: TS DOA}. TSL framework enables the model to assimilate information from the video modality alongside the guidance and regularization imparted by the audio teacher model, which ensures a balanced contribution of multi-modal features.

\begin{align}
\label{eq: TS-SELD} \mathcal{L}_{\mathrm{SELD}} = & \gamma_{1} \times \mathcal{L}_{\mathrm{SELD}}^{\mathrm{S}}+\gamma_{2} \times \mathcal{L}_{\mathrm{SELD}}^{\mathrm{TS}} \\
\label{eq: TS SED} \mathcal{L}_{\mathrm{SED}}^{\mathrm{TS}} = & \frac{1}{K N}\sum_{k, n}\left[\hat{y}_{k, n}^{\mathrm{T}} \log \frac{\hat{y}_{k, n}^{\mathrm{T}}}{\hat{y}_{k, n}^{\mathrm{S}}} \right. \\
& \left.+\left(1-\hat{y}_{k, n}^{\mathrm{T}}\right) \log \frac{\left(1-\hat{y}_{k, n}^{\mathrm{T}}\right)}{\left(1-\hat{y}_{k, n}^{\mathrm{S}}\right)} \right] \nonumber \\ 
\label{eq: TS DOA} \mathcal{L}_{\mathrm{DOA}}^{\mathrm{TS}} = & \frac{1}{K N}\sum_{k, n}\Big\|\left(\hat{\mathbf{o}}_{k, n}^{\mathrm{S}}-\hat{\mathbf{o}}_{k, n}^{\mathrm{T}}\right)\hat{y}_{k, n}^{\mathrm{T}}\Big\|^{2}
\end{align}where $\gamma_{1}=1$ and $\gamma_{2}=0.5$ are weighting factors for the student loss and teacher-regularized loss, separately. $\hat{y}_{k, n}^{\mathrm{S}}$ and $\hat{\mathbf{o}}_{k, n}^{\mathrm{S}}$ are the estimated active probabilities and DOA for the $n$-th sound event at the $k$-th frame in the student network, respectively. The parameters of the teacher model are fixed during the cross-modal learning.

\subsection{Video-guided decision fusion}
\label{ssec: fusion}
In the late stage, we employ a decision fusion approach to further uncover crucial information from video detection in order to enhance the model's localization precision. Video detection-based localization tends to yield more accurate results compared to network predictions, especially for cases involving rapid movements of sound sources, due to potential drift and fluctuation. Therefore, we propose a video-guided decision fusion rule to rectify inaccurate DOA estimations using visual cues in the following three steps: 

Firstly, the human keypoints detected in Section \ref{ssec: av model} can be associated with specific sound classes. Specifically, mouths are associated with male speech, female speech, clapping and laughter classes, left and right hands are associated with water tap class, and left and right feet are associated with walk class. Other sound event classes are excluded due to the lack of corresponding visual objects (e.g., Knock) or poor detection performance (e.g., Door). Secondly, given a Cartesian vector DOA estimation $\hat{\mathbf{o}}=\left(\hat{a}, \hat{b}, \hat{c}\right)$ of a sound event at $t$-th frame predicted by model and all coordinates of human keypoints $\mathbf{V} = \left\{\mathbf{v}_{1}, \mathbf{v}_{2}, \ldots, \mathbf{v}_{p}\right\}$ detected from the current video frame, we can compute the angular distances \cite{politis2020overview} between them, as defined below:
\begin{equation}
    d\left(\hat{\mathbf{o}}, \mathbf{v}_{i}\right)=\arccos \left(\frac{\left \langle \hat{\mathbf{o}},  \mathbf{v}_{i}\right \rangle}{\left\|\hat{\mathbf{o}}\right\|\left\|\mathbf{v}_{i}\right\|}\right)
\end{equation}where $\mathbf{v}_{i}=\left(a_{i},b_{i},c_{i}\right), i\in\{1,\ldots,p\}$, and $p$ is the number of detected keypoints. Thirdly, we select the keypoint candidate $\hat{\mathbf{v}}$ with the smallest angular distance $\hat{d}$. If $\hat{d}$ is less than a pre-defined threshold $\sigma = 30^{\circ}$, we consider the video detection outcomes to be more accurate and replace $\hat{\mathbf{o}}$ with $\hat{\mathbf{v}}$ as the final DOA estimation. Utilizing video-guided decision fusion, the localization accuracy and precision of the network can be further improved.

\section{EXPERIMENTS}
\label{sec:exp}
\subsection{Experimental setup}
\label{ssec: setup}
We evaluate the SELD task on the official development set of the Sony-Tau Realistic Spatial Soundscapes 2023 (STARSS23) dataset \cite{Shimada2023starss23_arxiv} recorded in realistic spatial soundscapes in DCASE 2023 Challenge. The development set of STARSS23 is divided into a training part (dev-set-train) about 3.83 hours and a testing part (dev-set-test) about 3.22 hours. The labels of the official evaluation set have not been released, and our results on the evaluation set are available on the DCASE 2023 Task3 Challenge results page\footnote{https://dcase.community/challenge2023/task-sound-event-localization-and-detection-evaluated-in-real-spatial-sound-scenes-results\label{dcase2023}}. The basic dataset contains a total of 13 sound event classes. The audio recordings are collected in a 4-channel spatial format with a 24 kHz sampling rate. The video data are captured using a 360$^\circ$ camera with a resolution of 1920 $\times$ 960 at 29.97 frames per second. The details of audio-visual feature extraction and fusion are discussed in Section \ref{ssec: ao model} and Section \ref{ssec: av model}. In the DCASE 2023 SELD task, a convolutional recurrent neural network (CRNN) is utilized as the audio-visual baseline structure. Our main model architecture is ResNet-Conformer consisting of 8 attention heads \cite{niu2023experimental}, and the dimensions of the input, key and value vectors are set to 256, 32 and 32, respectively. Adam \cite{kingma2014adam} is adopted as the optimizer. The tri-stage learning rate scheduler \cite{park19e_interspeech} is used with an upper limit of 0.001 without TSL and 0.0001 with TSL, respectively. We evaluate all methods with SELD$_{score}$ \cite{mesaros2019joint}, as defined below:
\begin{equation}\label{eqseld}
\text{SELD}_{score} = \frac{1}{4}[\emph{ER}_{20^{\text{o}}}+(1-\emph{F}_{20^{\text{o}}})+\emph{LE}_{\text{CD}}^{'}+(1-\emph{LR}_{\text{CD}})]
\end{equation}where $\emph{ER}_{20^{\text{o}}}$ and $\emph{F}_{20^{\text{o}}}$ represent location-dependent error rate and F-score when the spatial error is within $20^{\circ}$. $\emph{LE}_{\text{CD}}^{'}=\emph{LE}_{\text{CD}}/\pi$, where $\emph{LE}_{\text{CD}}$ denotes the localization error between predictions and references of the same class. $\emph{LR}_{\text{CD}}$ is a simple localization recall metric.

\subsection{Experimental results and analysis}
\label{ssec: results}

\begin{table}[t]
	\renewcommand\arraystretch{1.25}
	\newcolumntype{L}[1]{>{\raggedright\arraybackslash}p{#1}}
	\newcolumntype{C}[1]{>{\centering\arraybackslash}p{#1}}
	\newcolumntype{R}[1]{>{\raggedleft\arraybackslash}p{#1}}
	\centering
	\caption{Performances comparison among different data augmentation methods on audio-only (AO) teacher model and audio-visual (AV) student model. CRNN is used as the official baseline. `DA1': AO Basic data, `DA2': AO Basic data + simulated data, `DA3': AO Basic + simulated data + ACS, `DA4': AO Basic + simulated data + ACS + mixup, `DA5': AV Basic data, `DA6': AV Basic data + ACS-VPS, `DA7': AV Basic data + ACS-VPS + mixup.}
	\label{tab: DA}\medskip
	\resizebox{8.5 cm}{!}{\begin{tabular}{c|c|c|c|c|c|c}
			\toprule[1 pt]
			  Model & Setup & ER$_{20^{\circ}}$ & F$_{20^{\circ}}$ & LE$\rm_{CD}$ & LR$\rm_{CD}$ & SELD$_{score}$ \\
			\midrule
            \multirow{5}{*}{Teacher} & CRNN+DA1 & 1.00 & 0.14 & 60.00$^{\circ}$ & 0.33 & 0.72 \\
            & RC+DA1 & 0.75 & 0.17 & 39.82$^{\circ}$ & 0.38 & 0.61 (15.3$\%\downarrow$)\\
            & RC+DA2 & 0.56 & 0.42 & 18.65$^{\circ}$ & 0.67 & 0.39 (45.8$\%\downarrow$)\\
            & RC+DA3 & 0.45 & 0.55 & 14.28$^{\circ}$ & 0.66 & 0.33 (54.2$\%\downarrow$)\\
            & RC+DA4 & 0.42 & 0.57 & 14.30$^{\circ}$ & 0.67 & 0.31 (56.9$\%\downarrow$)\\
            \midrule
            \multirow{4}{*}{Student} & CRNN+DA5 & 1.07 & 0.14 & 48.00$^{\circ}$ & 0.36 & 0.71 \\
    	      & RC+DA5 & 0.76 & 0.18 & 34.13$^{\circ}$ & 0.42 & 0.59 (16.9$\%\downarrow$)\\
            & RC+DA6 & 0.57 & 0.36 & 18.82$^{\circ}$ & 0.51 & 0.45 (36.6$\%\downarrow$)\\
            & RC+DA7 & 0.53 & 0.49 & 15.80$^{\circ}$ & 0.64 & 0.37 (47.9$\%\downarrow$)\\
            \bottomrule[1 pt]
	\end{tabular}}
\end{table}

\subsubsection{Experiments on audio-visual data augmentation}
\label{sssc: DA}
We evaluate the effectiveness of various audio-visual data augmentation methods on teacher and student models trained from scratch. As described in Table \ref{tab: DA}, we first apply our self-developed RC network to the basic training data and its good contextual and global modeling capabilities produce 15.3$\%$ and 16.9$\%$ improvements on the AO teacher and AV student models relative to the official CRNN baselines, respectively. The early audio-visual feature fusion combines the acoustic and visual cues to favor a 0.02 improvement in SELD scores for the student model relative to the teacher model. Subsequently, we step by step employ different data augmentation techniques on the teacher and student models, respectively. In teacher model training, we continually add simulation data, ACS method and mixup augmentation on top of the official basic data, leading to four different audio data configurations, identified as `DA1', `DA2', `DA3' and `DA4'. The injection of substantial external data and effective data augmentation techniques progressively enhance the model's robustness, however, it should be noted that the simulated audio cannot be aligned with the video data, making the student model lose considerable external data compared to the teacher model. Therefore, we initially apply the proposed audio-visual joint data augmentation by simultaneously performing VPS with ACS augmentation, expanding the duration of the audio-visual data from 3.83 hours to 30.64 hours, represented as `DA6'. The ACS-VPS method significantly advances the student model with a 36.6$\%$ SELD improvement, since the ACS-VPS method creates more diverse spatial positions through rotation and flipping without disturbing the realism and continuity of the recorded video pixels and reverberation conditions and overlapping segments in multi-channel audio. Then, applying the mixup method, denoted as `DA7', at the input layer further enhances the model's performance. Not only in our experimental phenomena but also in the official baseline result, AV models outperform AO models slightly due to redundancy and obstacles in extracting video cues, which motivates us to further enhance the student model's capability by fusing audio and visual information.

\begin{table}[t]
	\renewcommand\arraystretch{1.25}
	\newcolumntype{L}[1]{>{\raggedright\arraybackslash}p{#1}}
	\newcolumntype{C}[1]{>{\centering\arraybackslash}p{#1}}
	\newcolumntype{R}[1]{>{\raggedleft\arraybackslash}p{#1}}
	\centering
	\caption{Performances comparison on TSL using AO teacher model `T' to instruct AV student model `S' with various configurations. `T1': RC + DA2, `T2': RC + DA3, `T3': RC + DA4, `T4': CRNN + DA4, `S1': RC + DA5, `S2': RC + DA6, `S3': RC + DA7.}
	\label{tab: TSL}\medskip
	\resizebox{8.5 cm}{!}{\begin{tabular}{c|c|c|c|c|c}
			\toprule[1 pt]
			  Model & ER$_{20^{\circ}}$ & F$_{20^{\circ}}$ & LE$\rm_{CD}$ & LR$\rm_{CD}$ & SELD$_{score}$ \\
			\midrule
            T1-S1 & 0.64 & 0.28 & 33.93$^{\circ}$ & 0.57 & 0.49 (31.0$\%\downarrow$)\\
            T2-S1 & 0.51 & 0.39 & 28.37$^{\circ}$ & 0.58 & 0.42 (40.8$\%\downarrow$)\\
            T3-S1 & 0.51 & 0.41 & 27.48$^{\circ}$ & 0.64 & 0.40 (43.7$\%\downarrow$)\\
            T3-S2 & 0.43 & 0.55 & 14.42$^{\circ}$ & 0.68 & 0.32 (54.9$\%\downarrow$)\\
            T3-S3 & 0.41 & 0.59 & 14.10$^{\circ}$ & 0.73 & 0.29 (59.2$\%\downarrow$)\\
            \midrule
            T4 & 0.53 & 0.42 & 18.33$^{\circ}$ & 0.60 & 0.40 \\ 
            T4-S3 & 0.47 & 0.49 & 15.91$^{\circ}$ & 0.63 &  0.36 (49.3$\%\downarrow$)\\
            \bottomrule[1 pt]
	\end{tabular}}
\end{table}

\subsubsection{Experiments on teacher-student learning}
\label{sssc: TSL}
Based on the teacher and student models trained as described above, we conduct experiments with cross-modal teacher-student learning. Following the TSL framework we proposed in Section \ref{ssec: cross-modal tsl}, the learning process of student models can be instructed by various teacher models. As illustrated in Table \ref{tab: TSL}, given the student model `S1' with the limited basic audio-visual dataset `DA1', different teacher models `T1', `T2' and `T3' are adopted first. As the guidance from teacher models grows, the SELD performance of the student models consistently improves from 0.49 to 0.40 (as depicted in the first three rows of Table \ref{tab: TSL}). Consequently, we fix the optimal audio model with `DA4' in Table \ref{tab: DA} as the teacher and conduct ablation experiments to assess the student model's performance (as exhibited in the third to fifth rows of Table \ref{tab: TSL}). By incorporating data augmentation techniques into the student model incrementally, TSL demonstrates a notable enhancement of 0.11 in SELD score on the AV student model, culminating in a 59.2$\%$ improvement relative to the official baseline. Additionally, in order to explore the generalization of our proposed TSL framework, we train an AO teacher model with the largest dataset 'DA4' based on the official CRNN network, which is denoted as 'T4'. Then the teacher model 'T4' is used to guide the AV student model with the RC network. Analogously, our TSL framework showcases a development of 49.3$\%$. Cross-modal TSL contributes to utilizing rich information from audio data and balancing original student loss $\mathcal{L}_{\mathrm{SELD}}^{\mathrm{S}}$ with the teacher-instructed regularized loss $\mathcal{L}_{\mathrm{SELD}}^{\mathrm{TS}}$, which helps alleviate the overfitting problem on low-resource realistic audio-visual dataset. Moreover, TSL framework is robust to different model architectures.

\begin{table}[t]
	\renewcommand\arraystretch{1.25}
	\newcolumntype{L}[1]{>{\raggedright\arraybackslash}p{#1}}
	\newcolumntype{C}[1]{>{\centering\arraybackslash}p{#1}}
	\newcolumntype{R}[1]{>{\raggedleft\arraybackslash}p{#1}}
	\centering
	\caption{Performances comparison for video-guided decision fusion (denoted as `VGDF') on teacher and student models.}
	\label{tab: VGDF}\medskip
	\resizebox{8.5 cm}{!}{\begin{tabular}{c|c|c|c|c|c|c}
			\toprule[1 pt]
			Model & VGDF & ER$_{20^{\circ}}$ & F$_{20^{\circ}}$ & LE$\rm_{CD}$ & LR$\rm_{CD}$ & SELD$_{score}$ \\
            \midrule
			\multirow{2}{*}{Teacher} & - & 0.42 & 0.57 & 14.30$^{\circ}$ & 0.67 & 0.31 \\
			                                & $\checkmark$ & 0.40 & 0.58 &12.64 $^{\circ}$ & 0.67 & 0.30 \\
            \midrule
            \multirow{2}{*}{Student} & - & 0.41 & 0.59 & 14.10$^{\circ}$ & 0.73 & 0.29  \\
                                            & $\checkmark$ & 0.40 & 0.61 & 12.25$^{\circ}$ & 0.73 & 0.28  \\
			\bottomrule[1 pt]
	\end{tabular}}
\end{table}

\begin{figure}[t]
\centering
\includegraphics[width=0.48\textwidth]{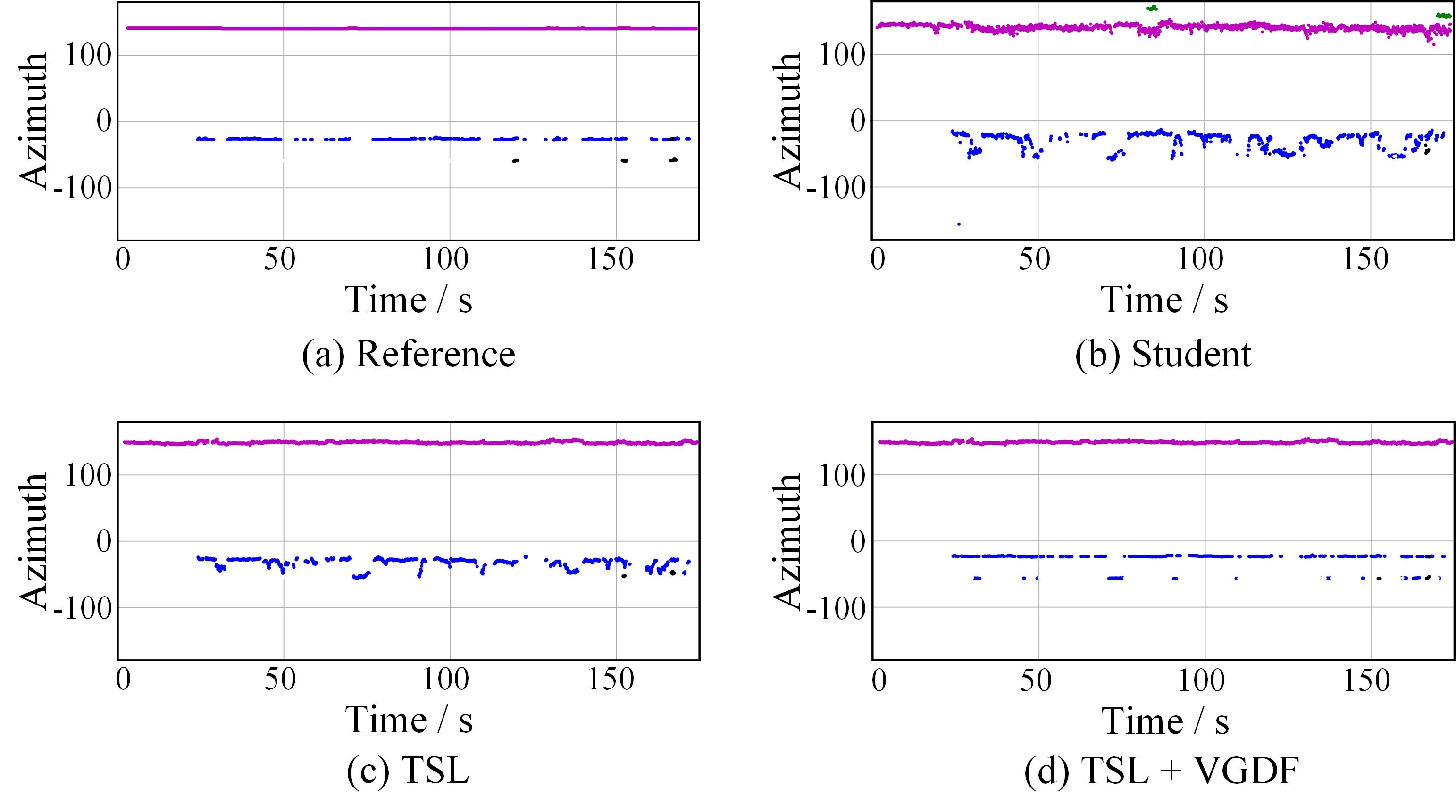}
\caption{The visualized comparison of azimuth DOA results of different methods, including student model trained with RC network and DA7 in Table \ref{tab: DA}, TSL model denoted as T3-S3 in Table \ref{tab: TSL} and TSL + VGDF method in Table \ref{tab: VGDF}. Different colors identify different sound event classes.}
\label{fig: visualize}
\end{figure}

\subsubsection{Experiments on video-guided decision fusion}
\label{sssc: VGDF}
At the late stage, we apply the video-guided decision fusion (VGDF) method to mine DOA information from video data on teacher and student models in Section \ref{sssc: TSL}, almost all of which demonstrate stable improvements in SELD performance. As shown in Table \ref{tab: VGDF}, we present the VGDF performance on the teacher model (refer to `RC+DA4' in Table \ref{tab: DA}) and the student model (refer to `T3-S3' in Table \ref{tab: TSL}). Upon closer inspection of metrics, it is evident that the VGDF method primarily optimizes the LE$\rm_{CD}$ metric, which aligns with our motivation for designing it. The localization error of the student model decreases from 14.10$^\circ$ to 12.25$^\circ$ thanks to the accurate video target detection algorithms and the matching rules we designed. Furthermore, the localization-dependent ER$_{20^{\circ}}$ and F$_{20^{\circ}}$ metrics are also optimized accordingly due to the correlation between the localization and detection metrics. The final single AV student model with VGDF achieves a SELD score of 0.28, which outperforms the second place in the DCASE 2023 Challenge by 15.2$\%$.

We visualize azimuth results of one recording in Fig. \ref{fig: visualize} to illustrate impacts of the proposed methods. Fig. \ref{fig: visualize} (b) is the result of the student model with AV joint augmentation, while TSL in Fig. \ref{fig: visualize} (c) gives precise SELD estimations, such as correcting \textit{Music} prediction (compared to the green line in Fig. \ref{fig: visualize} (b)). By further performing VGDF, Fig. \ref{fig: visualize} (d) exhibits the closest SELD performance to the reference in Fig. \ref{fig: visualize} (a) among all of the predictions.

\section{CONCLUSION}
\label{sec:conclusion}
We propose an audio-visual information fusion framework for SELD in low-resource realistic scenarios. By incorporating cross-modal TSL, multi-modal fusion and novel AV joint augmentation, the proposed framework demonstrates significant improvements and our submission to the SELD task of the DCASE 2023 Challenge won 1st place with model ensembles, standing out as the only one whose AV systems outperformed AO systems of all other teams in the Challenge. In the future, we will explore more cross-modal fusion strategies and inter-modal relationships to learn useful cues embedded in multi-modality data.

\section*{Acknowledgment}
This work was supported by the National Natural Science Foundation of China under Grant No. 62171427.

\bibliography{ref}
\bibliographystyle{IEEEbib}

\end{document}